\documentclass[5p,times,sort&compress]{elsarticle}

\usepackage{pgfplots}
\pgfplotsset{compat=newest}
\usetikzlibrary{patterns}

\usepackage{booktabs, tabularx, multirow}
\usepackage{amssymb}
\usepackage{amsmath} 
\usepackage{booktabs}
\usepackage{array}   
\usepackage{xcolor}
\usepackage{graphicx}
\usepackage{float}
\usepackage{rotating,lscape}
\usepackage{comment}
\usepackage{import}
\usepackage[hidelinks]{hyperref}
\usepackage[capitalise]{cleveref}
\usepackage{ragged2e}
\usepackage{microtype}

\hypersetup{
pdfinfo={
    Title={Introducing a Novel Systems Thinking approach inspired by STPA: Road Safety Intervention design case study},
    Author={Halima El Badaoui, Mariat James Elizebeth, Siddartha Khastgir, Paul Jennings},
}
}

\begin{document}

\begin{frontmatter}

\title{Introducing a Novel Systems Thinking approach inspired by STPA: Road Safety Intervention design case study}

\author[inst1]{Halima El Badaoui\corref{cor1}}
\ead{halima.el-badaoui@warwick.ac.uk}
\cortext[cor1]{Corresponding Author}
\author[inst1]{Siddartha Khastgir}
\author[inst1]{Mariat James Elizebeth}
\author[inst1]{Shufeng Chen}
\author[inst2]{Takuya Nakashima}
\author[inst1]{Paul Jennings}
\affiliation[inst1]{organization={WMG, University of Warwick},
addressline={6 Lord Bhattacharyya Way}, 
city={Coventry},
postcode={CV4 7AL}, 
state={West Midlands},
country={United Kingdom}
}
\affiliation[inst2]{organization={ Graduate School of Frontier Sciences, The University of Tokyo},
            addressline={ 5-1-5 Kashiwanoha }, 
            city={Kashiwa},
            postcode={277-8563}, 
            country={Japan}}

\begin{abstract}
According to the latest provisional statistics released by the UK Department for Transport, Great Britain recorded 1,633 road deaths in 2024, representing a slight increase from 2023 and raising concerns about safety progress, which indicates that preventable fatalities remain a challenge. The deployment of advanced mobility systems, even certified and safety-assessed, is not sufficient to deliver improved safety outcomes, and existing road infrastructure is not sufficiently equipped to prevent severe collisions. Successful application of the ``Safe System'' approach demands systems thinking in an integrated and holistic manner, encompassing all aspects of road safety. This paper argues that road safety must be managed as a complex socio-technical system where risk evolves dynamically and must be continuously monitored. To address these safety gaps, we propose a systems thinking approach that identifies factors contributing to fatal outcomes and mitigates them. The framework consists of four steps: 1) List stakeholders who influence road safety, 2) Model the interactions between these stakeholders, 3) List assumptions that might be identified as factors for fatalities, and 4) Monitor these assumptions throughout the system lifecycle. The approach is applied to the United Kingdom (UK) road network to demonstrate feasibility. The study provides actionable guidance and new KPIs categories for stakeholders to implement road safety monitoring and eliminate any unreasonable road safety risks.


\end{abstract}

%




\end{frontmatter}


\section{Introduction}

Reducing fatalities and increasing the survival rate in road transport remains a global challenge, despite the rapid deployment of advanced safety technologies such as Advanced Driver Assistance Systems (ADAS) and Automated Driving Systems (ADS). While these technological advancements are widely promoted for their potential to improve road safety, a reduction in serious and fatal crashes has not yet materialised. Achieving meaningful and measurable safety gain requires a comprehensive approach that integrates public perception and infrastructure readiness into every stage of system design and throughout the operational life cycle \cite{wang2013effect, Hollnagel2014}. Neglecting these essential interactions between people, technology and infrastructure not only undermines the potential benefits of automation but may also introduce new risks that might increase the likelihood of serious or fatal incidents \cite{Stanton2017, OECD2023}. The safe deployment of these technologies depends on social acceptance and the adaptability of existing road infrastructure, factors often overlooked in technical analyses.
Statistical evidence reinforces the necessity of a systemic approach. The World Health Organisation reports that 1.9 million people die each year from road traffic injuries and millions suffer from disability \cite{world2019global}.

Persistence or rising Killed or Seriously Injured (KSI) statistics worldwide demonstrate that current strategies, often built around retrospective crash data, are not achieving the targeted reductions in harm \cite{DfT2024,world2019global}. Multiple critical factors from road geometry, infrastructure quality and driver behaviour impact the severity of road crashes and their outcomes\cite{wang2013effect}. If strategies and action plans fail to address the combined and evolving effects of these factors, especially as new technologies are introduced, preventable losses will continue to occur \cite{wegman2017future}.
Traditional analytical methods such as Bow-Tie Analysis, Fault Tree Analysis (FTA), Event Tree Analysis (ETA), Failure Mode and Effects Analysis (FMEA), and Hazard and Operability Analysis (HAZOP), alongside road-specific approaches such as blackspot analysis\cite{elvik2007state}, and crash-based statistical models, remain reactive and limited in scope. They mainly focus on historical collision data and correlations, rather than the underlying systemic causes of risk \cite{elvik2009handbook,Reason1990,Heinrich1931}.
These methods tend to assume that risk is static and predictable and overlook feedback loops and emergent properties arising from stakeholder interactions.
The current stagnation in fatality and injury reduction signals the need to shift from retrospective crash data analysis toward continuous system-level risk monitoring \cite{international2016zero,UNDecade2021,ITFSafeSystem2016}.

Recent developments in safety science, particularly the System-Theoretic Accident Model and Processes (STAMP) perspective, have helped shift the focus from linear accident causation to understanding accidents as emergent properties of complex socio-technical systems\cite{Leveson2012,Thomas2015}. STAMP-based method, such as System-Theoretic Process Analysis (STPA), incorporates system thinking to enable hazard identification and the definition of safety constraints across technical, human and organisational factors. However, while systems thinking can reveal what might go wrong, it does not guarantee that identified hazards and assumptions will remain valid once systems are deployed \cite{INCOSE2023}. The challenge of maintaining the validity of assumptions throughout the entire lifecycle, particularly as road systems adapt to technological, behavioural and policy changes, remains insufficiently addressed.

\subsection{Literature Review}

Recent national and international road safety reports highlight that the persistent issue of reducing injury rates remains unresolved. In the United Kingdom, recent statistics published by the Department for Transport (DfT) report that "\emph{There has been no significant reduction in serious injuries}". Despite years of investigation and investment in road safety programmes, the progress in reducing fatalities has stagnated.
In  \cite{bachani2025time},the authors further highlights a critical mismatch between global estimates of road safety performance and local data. This discrepancy creates uncertainty regarding what is actually being addressed by policymakers and researchers. which may help explain the lack of improvement in road safety outcomes. Traditional approaches focus mainly on road safety behaviour, treating road safety as a transport issue.
Consequently, the authors advocate adopting a systems approach to road safety, which considers it as an emergent property of a complex socio-technical system, and systems thinking offers a suitable framework for this purpose, as it enables the interactions between system components, including road users, government, infrastructure, and vehicle manufacturers.

Building upon this foundation, \cite{mooren2024systems} explores the historical development of systems thinking and investigating transport and industry safety management to gain insight into how best to use the experiences in general road safety management. Similarly, a study that addresses the challenges associated with the Safe System approach and illustrates how inadequate road system design can hinder safe road user behaviour is presented in \cite{williamson2021we}. Numerous studies conducted by Salmon and his colleagues have demonstrated that Systems Theory, originally derived from workplace safety, can uncover the complex web of interacting factors across multiple levels of the road transport system that contribute to crashes \cite{salmon2013crash}\cite{salmon2020big}. An article that examines the ‘drift into failure’ approach for road safety efforts, which argues that road transport systems do currently display characteristics of complex systems in drift, and that a greater understanding of complexity theory-based models will support improved road safety efforts, is presented in \cite{salmon2012road}.

From a system theoretic safety perspective, STPA has been applied to Austria's road safety to expose gaps in the road safety inspection (RSI) process, especially the limited treatment of human, vehicle, and road interactions. The authors identify three main areas for improvement: 1) extend RSI beyond the TEN-T(Trans-European network) network to Non-TEN-T roads, 2) broaden the scope to include human factors, vehicle dynamics, and road user interactions, and 3) apply STPA within the overall road safety management process.Their key findings reveal that the current RSI practice in Austria is reactive rather than proactive, inspection boundaries are too narrow, excluding human, vehicle, and road interactions, and feedback loops are missing. These findings highlight a critical limitation: the absence of closed-loop monitoring to ensure implementation of recommendations, along with gaps in procedures and regulatory support needed to implement the proposed improvements\cite{kraut2021stpa}.
While systems approaches and safety analysis like STPA provides valuable insights and highlight area of improvement, there remains a need for a new framework that can deliver ongoing, operational assurance, through the continuous monitoring of underlying system assumptions.

As applied currently, STPA remains primarily a design hazard analysis and does not prescribe a run-time monitoring framework. While STPA's prominent role at design time is widely recognised, establishing continuous, operational monitoring requires additional assurance steps. As Leveson notes \cite{leveson2013systems}, the concept of assumption-based leading indicators provides guidance to identify and monitor key assumptions across the entire lifecycle. Yet, it does not specify metrics for every assumption, nor does it establish a mechanism for continuous, operational monitoring of assumptions.
\cite{dokas2013ewasap} addresses this gap by proposing Early Warning Sign Analysis based on the STPA (EWaSAP), an extension to STPA that is built around perceivable signs. EWaSAP helps analysts identify observable early warning signs and supports the configuration of sensors and awareness actions so that such signals are transmitted and monitored effectively. In their case study at a drinking water treatment works, EWaSAP started from an established STPA, first identifying signs indicating violations, next identifying systems with sensors capable of perceiving those signs, and finally stakeholders. This process, however, produced a large number of early warning signs, and the author acknowledges the complexity of managing such a multitude of early warning signs. EWaSAP  as an important extension to STPA that emphasises perceivable signs, non-perceivable assumptions may require complementary assurance mechanisms(e.g., audit, pass/ fail rate) to keep the monitoring burden tractable.

\subsection{Paper Contribution and Novelty}
\begin{figure*}[tb]
    \centering
    \includegraphics[width=\linewidth]{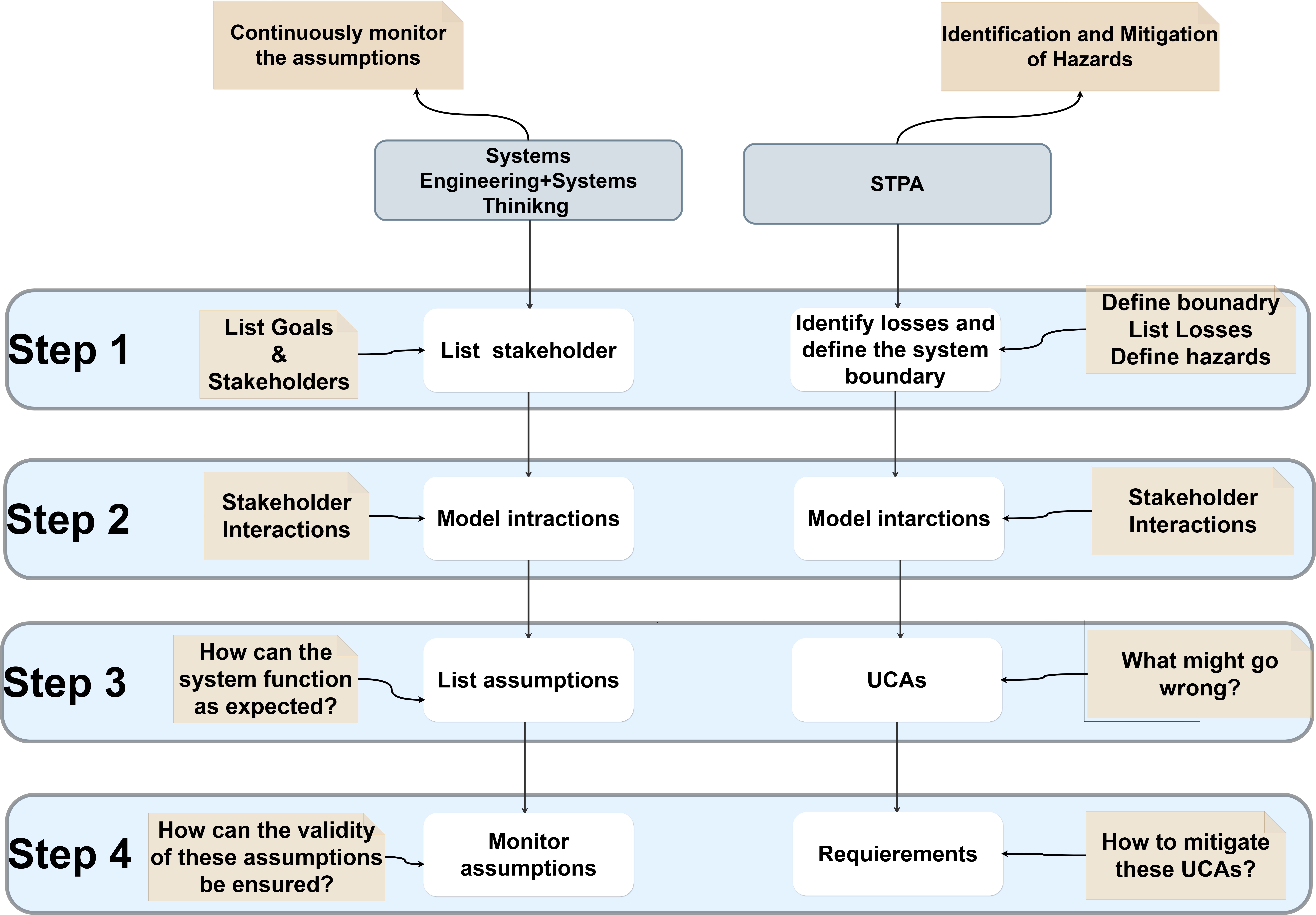}
    \caption{Overview of the proposed approach inspired by the STPA}
    \label{overview of prop app by stpa}
\end{figure*}

To address these limitations, this paper proposes the integration of systems thinking with systems engineering to enable continuous safety assurance across the entire lifecycle. Systems engineering introduces lifecycle governance, configuration management, verification and validation and traceability from hazards identification to deployment and operation \cite{INCOSE2023}. Unlike traditional safety analysis, system engineering principles provide robust mechanisms to monitor system behaviour and continuously validate the assumptions on which systems, policies, and safety decisions are based, ensuring they remain valid as technologies, regulations, and user behaviour evolve.
The novelty of the proposed approach is a structured framework to monitor both perceivable assumptions which can be directly observed or measure through operational indicators (e.g., calibration range, sensor accuracy...) and non-perceivable assumptions, that cannot be directly observed and therefore require indirect verification or periodic validation through audits or documentation (e.g., competence currency, policy acknowledgement...), enabling runtime traceability from the stakeholder through the assumptions to the leading indicator, action-owner, and corrective action if an assumption is invalid.
This approach introduces a detection mechanism that can trace changes in stakeholders' roles or responsibilities at run time, addressing an operational gap not covered by EWaSAP. Unlike EWaSAP, it does not require a pre-established STPA artefact, making it especially suitable when the study objective is assurance rather than hazards identification. This delivers a practical, explicit framework to ensure assumption validity, precisely where conducting a full STPA analysis can be unnecessarily lengthy or outside project scope.
It is essential to note that this approach does not aim to replace STPA, but rather focuses on safety assurance in an operational context by monitoring assumptions, rather than conducting hazard analyses of socio-technical systems.
The proposed approach retains STAMP's system thinking foundation to understand and analyse complex systems through the examination of interactions and mutual influences among their social and technical components, and combines it with system engineering disciplines. Unlike STPA, which begins by identifying losses to be mitigated, this method starts by listing stakeholders and defining goals to be achieved.
The second step is similar to STPA's control structure modelling. The third step lists assumptions based on each stakeholder's responsibilities and inputs, i.e., how assumptions are formed from received control actions and feedback from other stakeholders. In contrast, STPA's third step focuses on identifying unsafe control actions. The fourth step of the proposed approach centres on ensuring and tracking the validity of each assumption by identifying specific leading indicators, metrics for continuous monitoring and corrective actions to be initiated upon detection of violations. This aligns with STPA's step four, which defines requirements to mitigate unsafe control actions. Thus, the approach adopts system thinking from STPA but extends it to provide a new method for monitoring assumptions and managing safety across the entire system lifecycle. \Cref{overview of prop app by stpa} presents the mapping between the proposed approach and STPA, emphasising their structural and methodological differences.
Currently, the approach does not require a pre-established STPA to begin monitoring; it can link to STPA results when available, but can also operate independently when the scope is operational assurance.

Accordingly, we propose a four-step system engineering approach inspired by STAMP base STPA: 
\begin{itemize}

    \item \textbf{Step 1}: Identification of stakeholders (hardware components, software components, organisational, etc.),
    \item \textbf{Step 2}: Creation of a nested control structure for identified stakeholders,
    \item \textbf{Step 3}: Listing of assumptions, and step,
    \item \textbf{Step 4}: Monitoring of assumptions through the systems lifecycle.

\end{itemize}
This will answer the following research questions: 
\begin{itemize}
    \item \textbf{RQ1}- How to ensure the validity of assumptions through the entire lifecycle?
    \item \textbf{RQ2}- How can systems engineering incorporating systems thinking be effectively combined to enable an injury reduction and an increase in survival rates?

\end{itemize}

By continuously monitoring assumptions, this approach aims to bridge the gap between hazard identification and post-deployment. This supports the global effort to reduce fatalities and increase survival rates by transforming road safety management from a reactive, data-driven process into a proactive, adaptive system.

\section{Methodology}

\begin{figure*}[tb]
    \centering
    \includegraphics[width=0.8\linewidth]{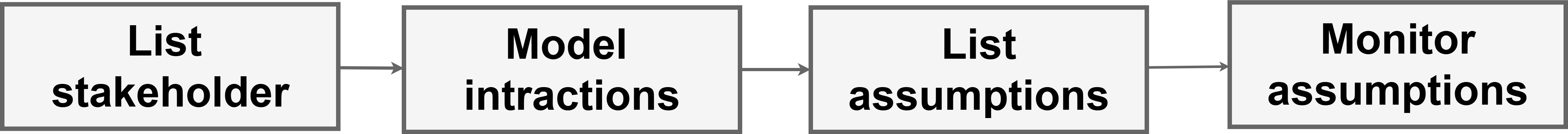}
    \caption{Methodology overview}
    \label{fig:SEmethodology}
\end{figure*}

\begin{figure}
    \centering
    \includegraphics[width=1\linewidth]{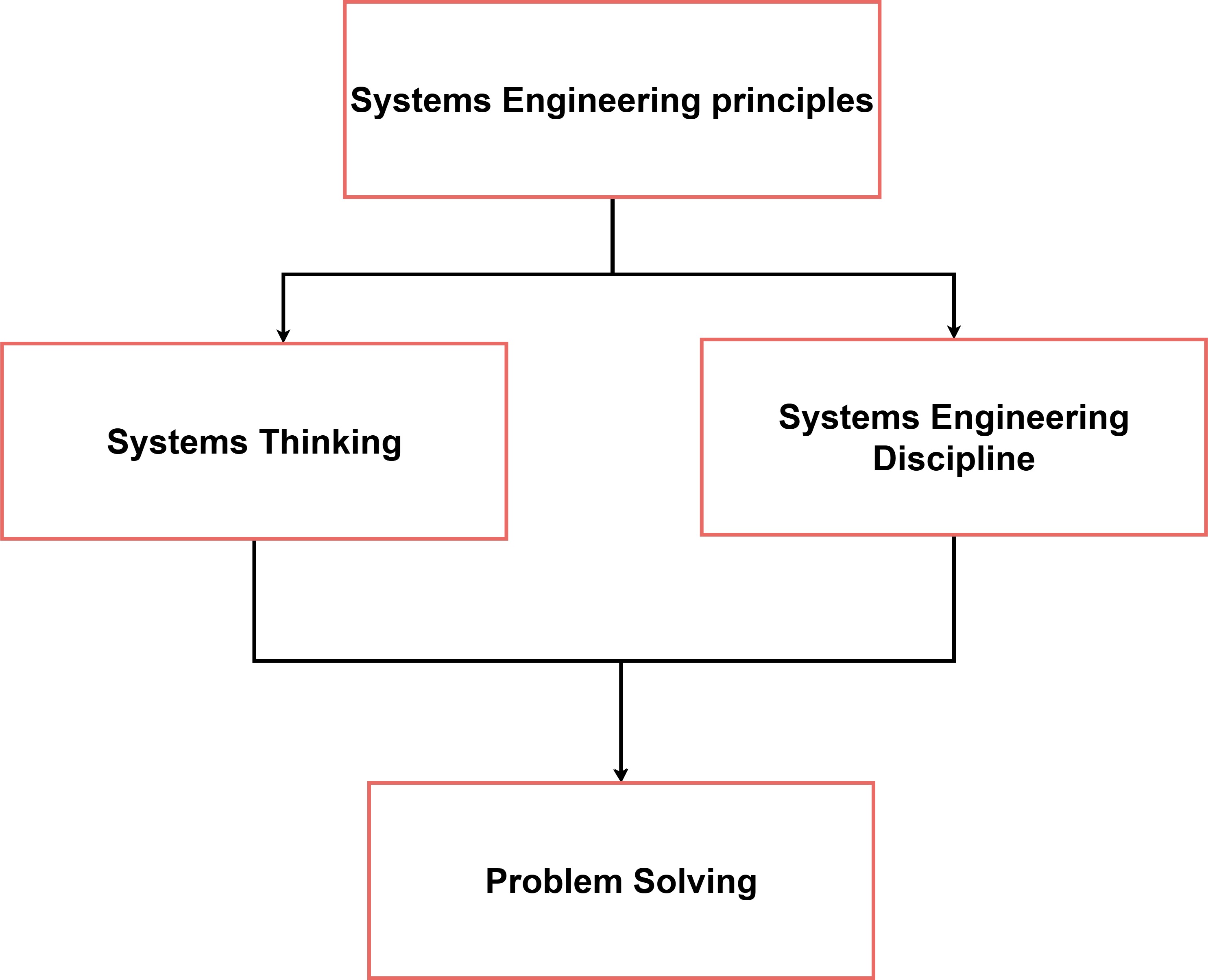}
    \caption{SE Principles}
    \label{fig:SE}
\end{figure}

To ensure that system safety is maintained consistently throughout the entire lifecycle, this paper introduces a structured approach based on four steps that integrate systems engineering principles, as systems thinking and systems engineering discipline are important foundations of system engineering principles, see \cref{fig:SE}. This integration enables the effective management of system complexity and lifecycle change, both of which are essential in modern socio-technical systems such as road transport. The proposed approach functions as a problem-solving framework that combines two foundational principles: systems thinking, which enables a holistic understanding of the system by examining its boundaries, components, interactions, and feedback loops; and systems engineering, which provides a disciplined process model to ensure that safety-related assumptions and requirements remain valid from system conception through deployment, operation, and retirement. \Cref{fig:SEmethodology} summarises the key steps of the proposed approach.

  \textbf{Identifying stakeholders:} This step aims to define the complete boundaries of the system under consideration. According to system engineering principles, a system cannot be adequately described and fully understood without specifying not only its internal components but also external interfaces and interrelationships \cite{benjamin2016system}. In system engineering, systems are generally considered open; their elements interact not only internally but also with the surrounding environment. Stakeholders, which include organisational, Humans, hardware, software, and infrastructure. These stakeholders collectively contribute to or influence the system lifecycle, including development, operational, maintenance, and retirement. Therefore, identifying stakeholders ensures all actors involved in or impacted by system safety and performance are captured within the system boundaries. The latter may vary depending on the focus of the analysis. This step provides a complete system definition and the basis required to define responsibilities and model interactions.
  As emphasised in the literature \cite{blanchard2004system}, the following four characteristics should be considered for a comprehensive system definition: 
  
  \begin{enumerate}
      \item Resource listing: A system consists of a complex combination of resources, including human, materials, organisations, and data.  
      
      \item Subsystem decomposition: Systems can be broken down into subsystems and components. It is essential to analyse both internal and external interactions.
      \item Purpose definition: a system must have a clearly defined purpose, and all included elements should contribute to fulfilling this purpose.
  \end{enumerate}
  To ensure completeness, stakeholders should be identified across all system lifecycle phases from multiple viewpoints, such as organisational, technical regulatory.

 \textbf{Identifying interactions between stakeholders:} After establishing the system boundary and listing all stakeholders involved, the next step is to model the hierarchical structure of interactions, aligning with systems engineering principles. This involves not only identifying authority paths but also operational relationships to define stakeholders' value.
 Stakeholders are positioned within this interaction hierarchy according to their responsibilities,  authority and impact on other stakeholders. 
The control structure must explicitly capture control actions and feedback, thereby forming a closed loop. By tracing these interactions at all hierarchical levels. Potential gaps or unsigned responsibilities can be spotted. it is also to define stakeholder value, to depict each stakeholder's contribution, responsibilities, and expectations in relation to system objectives.
This approach is consistent with the system-oriented view and network thinking advocated in systems engineering literature \cite{blanchard2004system}.

 \begin{itemize}
     \item \textbf{Closed loop control structure:} The control structure needed to form a closed loop decomposes into control action and feedback.
     \item \textbf{ Hierarchical and networked modelling:} systems are broken down into subsystems and components, and control relationships are traced at each level, facilitating detection of potential gaps.
     \item \textbf{Value and responsibilities:} stakeholders value the definition and allocation of responsibilities within the model.
 \end{itemize}

\textbf{Listing assumptions:} after modelling the interactions and having a better visualisation of stakeholders' interaction, this step aims to list assumptions that explicitly define each controller's process model. These assumptions articulate what each stakeholder presumes about the system state, the behaviour of other actors and the operating environment when issuing control action or interpreting feedback. Listing these assumptions is important to avoid discrepancies between assumed and actual system states \cite{haberfellner2019systems}. Assumptions may relate to system functionalities, response time, resource availability, environmental conditions, and a reliable feedback loop. Each assumption should be formulated clearly, specifying the conditions under which the system can achieve the expected system's purpose. 
 Assumptions can be categorised as either perceivable assumptions that can be continuously monitored through direct, observable indicators, or non-perceivable assumptions, which can lack telemetry (e.g., public perception, compliance rate, fraud...). Explicitly, these assumptions ensure traceability to enable their ongoing validation as the system and its environment evolve, for facilitating early detection.  

 To ensure a comprehensive list, a structured checklist guides stakeholders to record all assumptions that need monitoring for safe deployment, covering:

 \begin{itemize}
     \item Availability: Is the control action or feedback (CA/FB) available whenever it is required to support safe and effective operation?
     \item Timing: Does the CA/FB arrive soon enough to enable timely and correct decision-making within the control process?
     \item Accuracy: Is the information conveyed by the CA/FB accurate, reliable, and free from corruption or ambiguity?
     \item Authority: Is it clearly defined who has the authority to issue, approve, or override this CA/FB, especially in cases of conflict or uncertainty?
     \item Boundary: Within what operational or environmental boundaries does the stakeholder rely on this CA/FB being valid? 
 \end{itemize}

All responses to these guiding questions should be structured as follow:
 
 [\textcolor{blue}{Stakeholder}] assumes that [\textcolor{pink}{CA/FB}] [\textcolor{orange}{satisfies the relevant criteria availability, timing, accuracy, authority and boundary in a specific context}].


\textbf{Monitoring the assumptions:} To ensure that the assumptions listed in the previous step are subject to supervision as soon as they are defined, each assumption must be translated into a measurable term by identifying one or more leading indicators. This approach transforms assumptions from static documentation into dynamically supervised metrics of system safety. For every assumption, it is necessary to define early warning signals or metrics that indicate when an assumption may become invalid or is at risk of drift. These Leading Indicators may include, for example, calibration thresholds, data completeness, and update rates. Leading indicators should be chosen to allow for the prompt detection of emerging hazards before a loss occurs.
 For each monitored assumption, the following elements shall be specified: 
 \begin{itemize}

     \item \textbf{Leading Indicators}: The measurable variable that best reflects the validity of the assumption can be measurable leading indicators, or linked to evidence-based schedules (training records, audits, compliance inspections.
     \item \textbf{Thresholds and Criteria}: The quantitative or qualitative limits that define acceptable thresholds.
     \item \textbf{Corrective Action}: Actions to be initiated when thresholds are breached, including automatic safe state transition or operator notification.
     
 \end{itemize}

This approach enables continuous verification of the validity of the assumptions as the system or its operating environment evolves. It contributes to a more robust safety strategy by not only addressing identified hazards, as conventional safety analyses typically do, but also actively monitoring the underlying assumptions that support system design and behaviour.
As an outcome of this approach, all monitored elements are presented within the same operational dashboard to provide unified visibility and facilitate the early identification of assumption drift. It also enables traceability from stakeholders, assumptions, leading indicators, action-owner,  and corrective action.

\section{Case Study and Results}
\subsection{Identifying Stakeholders:}

In accordance with system engineering principles, a system cannot be fully defined without specifying both internal and external components. As part of the first step of our methodology, stakeholders must be identified to capture all actors that influence or are impacted by system behaviour during its lifecycle. This aligns with the four characteristics of comprehensive system definition: resource listing, hierarchy definition, subsystem decomposition, and purpose definition.

The objective of this case study is to examine the United Kingdom road transport safety system as a complex socio-technical system, to reduce road traffic fatalities and serious injuries while improving post-collision survival outcomes.A safe system approach is adopted to ensure a comprehensive coverage of all relevant stakeholders under five categories: safe road and roadsides, safe speeds, safe vehicles, safe road users, and post crash care.

\Cref{tab:stakeholders} presents the identified stakeholders within the context of the UK road transport safety system. These stakeholders contribute directly or indirectly to road transport safety performance, and the following steps.

\begin{table*}[tb]
\centering
\caption{Road Safety Stakeholders Categorised by Safe System Elements}
\label{tab:stakeholders}
\renewcommand{\arraystretch}{1.2}
\begin{tabular}{|p{2.6cm}|p{5.4cm}|p{6.2cm}|}
\hline
\textbf{Safe System Element} & \textbf{Stakeholders} & \textbf{Functional Role} \\
\hline

\multirow{6}{=}{\textbf{Safe Roads and Roadsides}} 
& Department for Transport (DfT) & National road safety strategy, funding, STATS19 ownership \\
& National Highways & Design, construction and operation of the Strategic Road Network \\
& Local Highway Authorities & Local road management, speed limits, engineering and maintenance \\
& ORR – Office of Rail \& Road & Safety and performance regulator of National Highways \\
& Research \& Standards bodies (RoSPA, BSI, TRL, academia) & Evidence generation, standards and guidance for infrastructure safety \\
& Research bodies & Evaluation of measures and publication of safety evidence \\
\hline

\multirow{5}{=}{\textbf{Safe Speeds}} 
& Home Office / NPCC Roads-Policing Lead & National roads-policing strategy and enforcement capability \\
& UK Police Forces – RPUs & Speed and traffic-law enforcement, collision investigation \\
& Fleet Operations Centre (FOC) & Telematics monitoring, speed compliance, incident alerts \\
& Freight \& Logistics Companies & Operational speed management policies \\
& Insurance & Risk event data, behavioural feedback, incentive mechanisms \\
\hline

\multirow{6}{=}{\textbf{Safe Vehicles}} 
& Vehicle Certification Agency (VCA) & Type approval of vehicles and components \\
& DVSA & MOT regulation, recalls oversight, vehicle standards \\
& Vehicle Manufacturers & Vehicle design, production, safety systems, recalls \\
& Vehicle Dealers & Recall execution, servicing, safety updates \\
& MOT Test Stations & Statutory roadworthiness inspections \\
& Maintenance Operations Centre (MOC) & Preventive maintenance, compliance and repair coordination \\
\hline

\multirow{7}{=}{\textbf{Safe Road Users}} 
& DVLA & Driver licensing, medical fitness, driver database \\
& Driving Schools / Training Providers & Driver training and competency development \\
& Road Users & Vehicle operation and behavioural compliance \\
& UK Police Forces – RPUs & Behavioural enforcement and investigation \\
& Freight \& Logistics Companies & Driver-hours control, safety policies \\
& Fleet Operations Centre (FOC) & Monitoring driver behaviour and fatigue \\
& Insurance & Behavioural risk profiling and feedback \\
\hline

\multirow{6}{=}{\textbf{Post-Crash Response \& Learning}} 
& NHS Ambulance Trusts & Pre-hospital emergency response \\
& Major Trauma Centres / NHS & Trauma treatment and survival audit \\
& Air Ambulance Services & Rapid medical evacuation \\
& Fire \& Rescue Services & Extrication and scene safety \\
& RSIB – Road Safety Investigation Branch & Independent investigation and systemic safety recommendations \\
& Insurance & Crash data analysis and loss information \\
\hline

\end{tabular}
\end{table*}

\subsection{Identifying interactions between stakeholders:}
Next step is to define the interactions between the identified stakeholders, in line with system engineering principles, which incorporate a systems thinking perspective to capture how safety performance emerges from the interaction between multiple stakeholders rather than from their isolated provided actions.
This perspective is highly needed in road safety, where outcomes such as collision prevention and post-crash survivability depend on coordinated actions across regulatory bodies, road operates, emergency services, vehicle manufacturers and road users.
These interactions include both authority paths and operational dependencies that define how control actions and feedback are modelled.
Visualising these interactions is fundamental to understanding how safety constraints are communicated, implemented across organisational levels.

\begin{figure*}[tb]
    \centering
    \includegraphics[width=0.97\textheight,height=1\linewidth,keepaspectratio,angle=90]{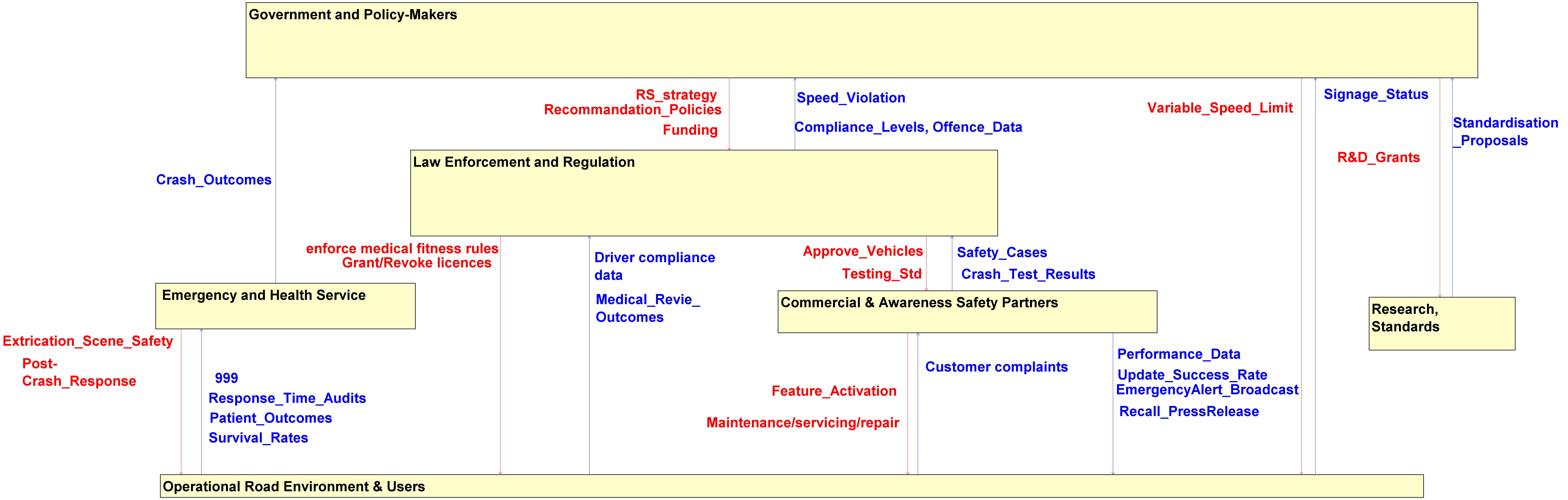}
    \caption{Control Structure of Road Safety Ecosystem}
    \label{fig: control structure}
\end{figure*}

However, due to the complexity of the UK road transport safety system and the large number of interacting actors, presenting every interaction would result in a dense structure, which would lead to voluminous content. Therefore, this study focuses on relevant safety-critical closed-loop control interactions. Specifically, those that directly impact the reduction of fatalities and increase the survival rate.
\begin{itemize}
    \item High-level control structure: provides a system-wide view of information and control flow across major stakeholders, presented in \cref{fig: control structure}.
\end{itemize}

\subsection{Listing assumptions}
Once the safety-relevant interactions have been modelled through closed loop control structures, in this step, an assumption list will be generated in order to ensure these closed loops are functioning as intended. These assumptions reflect what each stakeholder (controller) presumes about the system behaviour, resource availability, and reliability of feedback when issuing a control action. if these assumptions do not hold during operation, safety constraints may be violated, leading to critical losses.
 in \cref{tab:assumptions-ecall,tab:assumptions- Temporary Traffic Management (Work-Zones),tab:assumptions-Destination compliance,tab:assumptions-RSIaction} depicted examples of assumptions.
 \begin{quote}
\textit{Note.} The context underpinning the assumptions presented in this study was established using evidence from accident investigation reports, national and regional road safety performance reviews, transport authority datasets, and peer-reviewed research on crash causation and system performance.
\end{quote}




\textbf{Why do we need to monitor these assumptions ?}

These assumptions are safety-critical because delays in emergency response increase fatality risk.
It might be violated under several operating conditions, for example: 

\begin{itemize}
    \item If dispatch miscommunicates incident details, the wrong resources may arrive.
    \item If Fire \& Rescue delays extrication, medical care cannot begin.
    \item Road congestion or adverse weather conditions may increase ambulance response time beyond safe limits.
\end{itemize}


\begin{figure*}[tb]
    \centering
    \includegraphics[width=\linewidth]{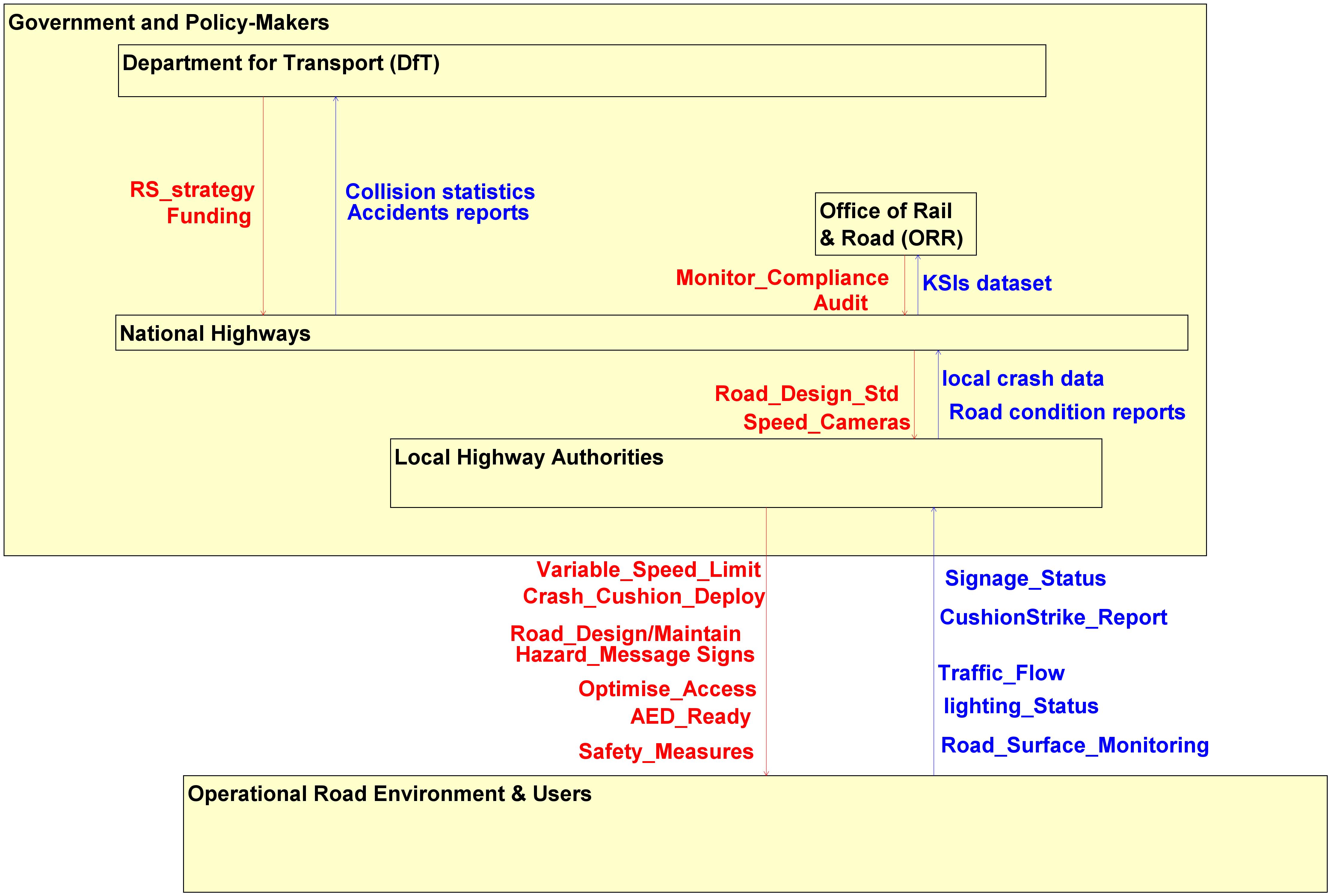}
    \caption{Variable Speed Limit}
    \label{fig:Variable Speed Limit}
\end{figure*}


\begin{itemize}
    \item Driver Confusion: Poorly displayed or inconsistent speed limits reduce compliance and create dangerous speed differentials between vehicles.
    \item System Failures: Undetected faults (e.g., non-functioning signs, algorithm errors) may persist, resulting in unsafe or misleading speed limit information.
\end{itemize}

\subsection{ Monitoring the assumptions}

Once we list all assumptions behind the closed loop in the control structure, each assumption needs to be translated into measurable terms to enable continuous monitoring. This step prevents assumptions from being violated and converts them from a statement into dynamic safety constraints. For each assumption, a set of leading indicators is defined, along with acceptable thresholds and predefined corrective actions to be taken if deviations are detected. \Cref{LI_CA} summarises the monitoring for the illustrative assumptions extracted from the control loop.


   
   


The current EU road safety policy as define by the European Commission focuses mainly on eight Key Performance Indicators ( KPIs ) for monitoring and benchmarking: speed, safety belt use, protective equipment, alcohol, distraction, vehicle safety Infrastructure, and post crash care. These KPIs are widely recognised and essential, but not sufficient for operational safety assurance \cite{EuropeanCommission2022KPIs}.  
Analsyisng the control structure, we mapped the control actions (CAs) and feedback (FB) to the existing categories, as presented in  \cref{existing_CS}. Additionally, new categories were created for the CAs and FB that did not correspond to any of the existing categories, as shown in \cref{Newexisting_CS}.

\begin{table*}[tb]
  \centering
  \caption{Assumptions about eCall MSD Delivery}
  \label{tab:assumptions-ecall}
  \setlength{\tabcolsep}{6pt}
  \begin{tabularx}{\textwidth}{p{2cm}Xlp{8cm}}
    \toprule
    \textbf{Assumption\_ID} & \textbf{From $\to$ To (stakeholders)} & \textbf{Category} & \textbf{Stakeholder assumes that …} \\
    \midrule
   $[A-1.1]$ & ADS Vehicle / OEM telematics $\to$ PSAP (999) & Availability & The eCall function is available and can be invoked whenever a qualifying collision occurs. \\\hline
    $[A-1.2]$ & ADS Vehicle / OEM telematics $\to$ PSAP (999) & Timing      & The Minimum Set of Data (MSD) is received in time to initiate an emergency response. \\\hline
    $[A-1.3]$ & ADS Vehicle / OEM telematics $\to$ PSAP (999) & Accuracy    & The MSD content is complete and correct and interpretable by PSAP systems. \\\hline
    $[A-1.4]$ & ADS Vehicle / OEM telematics $\to$ PSAP (999) & Authority   & eCall call setup and routing occur only through authorised 999 PSAP channels. \\\hline
    $[A-1.5]$ & ADS Vehicle / OEM telematics $\to$ PSAP (999) & Boundary    & eCall operation is valid within national 999 service coverage and supported communication technologies. \\
    \bottomrule
\end{tabularx}
\end{table*}

\begin{table*}[tb]
  \centering
  \caption{Assumptions about Destination compliance}
  \label{tab:assumptions-Destination compliance}
  \setlength{\tabcolsep}{6pt}
  \begin{tabularx}{\textwidth}{p{2cm}Xlp{8cm}}
    \toprule
    \textbf{Assumption\_ID} & \textbf{From $\to$ To (stakeholders)} & \textbf{Category} & \textbf{Stakeholder assumes that …} \\
    \midrule
    $[A-2.1]$ & Ambulance Service $\to$ Major Trauma Centre (MTC) / Trauma Unit (TU) & Availability & MTCs within the network are available to receive eligible patients. \\\hline
    $[A-2.2]$ & Ambulance Service $\to$ Major Trauma Centre (MTC) / Trauma Unit (TU) & Timing      & Patients meeting criteria are conveyed directly to an MTC within the required time window, unless immediate stabilisation at a TU is necessary. \\\hline
    $[A-2.3]$ & Ambulance Service $\to$ Major Trauma Centre (MTC) / Trauma Unit (TU) & Accuracy    & Field triage and pre-alert information used for destination decisions are accurate and up to date. \\\hline
    $[A-2.4]$ & Ambulance Service $\to$ Major Trauma Centre (MTC) / Trauma Unit (TU) & Authority   & The clinical authority for destination selection (crew lead/clinical desk) is clearly defined and applied. \\\hline
    $[A-2.5]$ & Ambulance Service $\to$ Major Trauma Centre (MTC) / Trauma Unit (TU) & Boundary    & Decisions comply with regional Major Trauma Network policies (e.g., drive-time thresholds, exclusion criteria). \\
    \addlinespace[2pt]
      \bottomrule
  \end{tabularx}
\end{table*}

\begin{table*}[tb]
  \centering
  \caption{Assumptions about Temporary Traffic Management (Work-Zones)}
  \label{tab:assumptions- Temporary Traffic Management (Work-Zones)}
  \setlength{\tabcolsep}{6pt}
  \begin{tabularx}{\textwidth}{p{2cm}Xlp{8cm}}
    \toprule
    \textbf{Assumption\_ID} & \textbf{From $\to$ To (stakeholders)} & \textbf{Category} & \textbf{Stakeholder assumes that …} \\
    \midrule
    $[A-3.1]$ & Road-works contractor / Road operator $\to$ Road users (via signs \& VMS) & Availability & Temporary signs and VMS are in place and functional whenever the work-zone is active. \\\hline
    $[A-3.2]$ & Road-works contractor / Road operator $\to$ Road users (via signs \& VMS) & Timing      & Signs/VMS are deployed before works commence and updated promptly when layouts change. \\\hline
    $[A-3.3]$ & Road-works contractor / Road operator $\to$ Road users (via signs \& VMS) & Accuracy    & Messages and layouts conform to the approved TTM plan and are legible and correct (Traffic Signs Manual). \\\hline
    $[A-3.4]$ & Road-works contractor / Road operator $\to$ Road users (via signs \& VMS) & Authority   & Any setup or change to the layout/message is approved by the designated TTM authority. \\\hline
    $[A-3.5]$ & Road-works contractor / Road operator $\to$ Road users (via signs \& VMS) & Boundary    & Deployment remains within the geometric and environmental limits specified in the approved plan . \\
    \addlinespace[2pt]
    \bottomrule
  \end{tabularx}
\end{table*}

\begin{table*}[tb]
  \centering
  \caption{Assumptions about RSI action}
  \label{tab:assumptions-RSIaction}
  \setlength{\tabcolsep}{6pt}
  \begin{tabularx}{\textwidth}{p{2cm}Xlp{8cm}}
    \toprule
    \textbf{Assumption\_ID} & \textbf{From $\to$ To (stakeholders)} & \textbf{Category} & \textbf{Stakeholder assumes that …} \\
    \midrule
    $[A-4.1]$ & Road Authority / ORR (inspector) $\to$ Road operator & Availability & An action-tracking system for RSI findings is available and in regular use. \\\hline
    $[A-4.2]$ & Road Authority / ORR (inspector) $\to$ Road operator & Timing      & Findings are tracked and closed within target deadlines defined by the safety-management procedure. \\\hline
    $[A-4.3]$ & Road Authority / ORR (inspector) $\to$ Road operator & Accuracy    & Records of findings, owners and evidence of closure are complete, correct and auditable. \\\hline
    $[A-4.4]$ & Road Authority / ORR (inspector) $\to$ Road operator & Authority   & Roles authorised to assign, approve and close actions are defined and enforced. \\\hline
    $[A-4.5]$ & Road Authority / ORR (inspector) $\to$ Road operator & Boundary    & Governance operates within the applicable road-infrastructure safety-management requirements. \\
    \bottomrule
  \end{tabularx}
\end{table*}

\begin{table*}[tb]
\centering
\caption{Mapping existing CAs and FB to the existing KPIs}
\label{existing_CS}
\renewcommand{\arraystretch}{1.2}
\begin{tabularx}{\textwidth}{p{4cm}XXX}
\hline
\textbf{EU KPI Domain} & \textbf{Examples CAs / FBs} & \textbf{Stakeholders involved} & \textbf{Examples of what is} \\
\hline
Speed (KPI1) & Speed\_Cameras,\newline  Variable\_Speed\_Limit,\newline  Enforcement\_Strategy & Police $\rightarrow$ RoadUsers $\rightarrow$ DfT & Vehicle speed compliance, enforcement coverage \\
\hline
Safety belt \& Protective equipment (KPIs 2–3) & Seatbelt\_Campaigns,\newline  Protective\_Equipment\_Campaigns & DfT / Police $\rightarrow$ Public & Seatbelt or helmet use monitoring \\
\hline
Alcohol (KPI4) & DrinkDriving\_Enforcement, BAC\_Test\_Results & Police $\rightarrow$ DfT / Courts & Share of drivers within BAC limit \\
\hline
Distraction (KPI5) & Distraction\_Campaigns,\newline  Mobile\_Phone\_Enforcement & Police $\rightarrow$ RoadUsers $\rightarrow$ DfT & Share of drivers not using handheld devices \\
\hline
Vehicle safety (KPI6) & Recall\_Execution,\newline Cyber\_Patch\_Management & OEM $\rightarrow$ VCA $\rightarrow$ RoadUsers & Type-approval safety, partial overlap (Baseline counts new cars only) \\
\hline
Infrastructure (KPI7) & Infrastructure\_Rating,\newline Winter\_Service,\newline Traffic\_Management\_Control & National Highways / Local Authorities & Share of road length meeting iRAP rating \\
\hline
Post-crash care (KPI8) & eCall\_Notification, PostCrash\_Response, AirAmbulance\_Dispatch, OnScene\_Care & ADSVehicle / PSAP / EMS / Hospitals & Time call$\rightarrow$arrival; trauma network capability \\
\hline
\end{tabularx}
\end{table*}

\begin{table*}[tb]
\centering
\caption{Mapping existing CAs and FB to new KPIs}
\label{Newexisting_CS}
\renewcommand{\arraystretch}{1.2}
\begin{tabularx}{\textwidth}{p{4cm}p{4.5cm}X}
\hline
\textbf{New KPI Category} & \textbf{Examples CAs / FBs } & \textbf{Purpose} \\
\hline
1. Lifecycle Governance & Audit\_Action\_Closure\newline Monitor\_Compliance\newline Governance\_Feedback\newline Funding\_Allocation & Tracks whether safety actions, audits, and policy decisions are completed and verified throughout lifecycle \\
\hline
2. Organisational Competence \& Role Assurance & Driver\_Testing\_Standards\newline Competence\_Certification\newline Training\_Update & Monitors that stakeholders remain competent and authorised to perform safety-critical roles \\
\hline
3. Data Integrity \& Reporting Chain Assurance & STATS19\_Data\newline Accident\_Reports\newline KSI\_Dataset\newline Data\_Exchange\_Policy & Validates reliability, timeliness, and completeness of data underpinning KPIs \\
\hline
4. Digital Safety \& Cybersecurity Assurance & OverTheAirSwUpdates\newline Cyber\_Patch\_Management & Ensures in-use software and connected systems remain safe and secure \\
\hline
5. Emergency Response Chain Assurance & PostCrash\_Response\newline AirAmbulance\_Dispatch\newline OnScene\_Care\newline Handover\_Protocol & Covers triage, destination, and quality of 999/eCall arrival \\
\hline
6. Infrastructure Operations \& Work-Zone Safety & Winter\_Service\newline Traffic\_Management\_Control\newline Temporary\_Orders\newline VMS\_Warnings & Manages transient risk from road works and maintenance \\
\hline
7. Fleet Operations \& Compliance & Preventive\_Maintenance\newline MOT\_Compliance\newline Telematics\_Monitoring\newline Enforcement\_Priorities & Monitors operational vehicle condition and operator compliance \\
\hline
8. Funding \& Strategic & Resource\_Allocation\newline Funding\newline Policy\_Performance & Ensures resources and budgets remain aligned with safety priorities \\
\hline
\end{tabularx}
\end{table*}

\begin{table*}[tb]
\caption{Example of leading indicators, their supervising owners, and corresponding corrective actions when an assumption is not valid.}
\label{LI_CA}
\centering
\begin{tabularx}{\textwidth}{lXlX}
\hline
\textbf{Assumption-ID} & \textbf{Leading indicator(s) (LI)} & \textbf{Indicator supervised by} & \textbf{Corrective action(s)} \\
\hline
[A-1.1] &
Rate of eligible crashes with successful MSD.\newline
PSAP message processing error rate. &
PSAP operations manager &
Training for operators\newline
Deploys AI-based software for data collection\newline
Regular audits of recorded calls \\
\hline
[A-2.2] &
Inter-facility transfer time &
Major trauma coordinator &
Air ambulance trigger\newline
Retaining protocols\newline
Optimise emergency road \\
\hline
[A-3.3] &
Rate of work-zone sites with approved TTM plan\newline
Audit non-conformity rate; VMS/signage uptime &
Work-zone safety manager &
Deploy crash cushions\newline
Repair / restore VMS\newline
Audits \\
\hline
[A-4.4] &
Rate of RSI actions closed on time\newline
Tracks backlog &
Road-authority safety &
Escalate overdue actions\newline
Adjust funding priorities\newline
Targeted reinspection\newline
Publish compliance notices \\
\hline
\end{tabularx}
\end{table*}

\clearpage
\section{Discussion}
This paper addressed a long-standing challenge in road safety: Despite the application of multiple safety analysis techniques, both proactive and reactive, fatality rates have not consistently decreased. Recent statistics show either a plateau or an increase in road deaths and serious injuries, indicating that existing safety assessment practices are insufficient to drive significant improvement. The persistent issue suggests that the problem may not lie in the lack of safety tools, but rather in what is being analysed and how. Traditional approaches focus heavily on accident data and hazard identification, which is important for that end, but only that might overlook a fundamental dimension of system behaviour: the assumption that the system design and operation are built upon.
This study contributes a novel perspective by shifting the safety focus from hazards alone to the assumptions embedded within the system control structure. Unlike conventional methods, this approach examines whether the system behaves as expected under real operating conditions and whether its presumed assumptions hold across the entire lifecycle. By integrating system engineering principles, which incorporate a system thinking perspective, the study enabled a holistic understanding of the road safety system across its lifecycle, capturing technical, organisational and operational interactions. 
The use of system thinking was essential, as it allowed the road safety system to be modelled not as an abstract set of actors but as a functional architecture of dependent stakeholders. Through control structure modelling, it became possible to visualise decision hierarchies, control action, and feedback flows across government, agencies, emergency services, infrastructure, commercial fleets, research bodies, and road users. The novelty of this work lies in making implicit system assumptions explicit and traceable. When assumptions remain undocumented, they are highly vulnerable to being overlooked, misunderstood or violated, leading to unsafe operating conditions without immediate detection. This response the (\textbf{RQ1}- How to ensure the validity of assumptions through the entire lifecycle?). As this ensure lifecycle validity by:
\begin{itemize}
    \item Maintaining an assumption documented linked to control loop.
    \item Defining leading indicators.
    \item Setting trigger thresholds for corrective action.
\end{itemize}
Documenting these assumptions was the critical first step; however, documentation alone is not sufficient. Safety cannot rely on static expectations; assumptions must be continuously validated as the system evolves. For this reason, the study introduced monitoring of assumptions using leading indicators. These Indicators provide early warning signals when reality begins to drift away from what the system presumes to be true. 

Our approach was applied to demonstrate how the explicit documentation, monitoring, and management of lifecycle assumptions can translate the static text or thought into tangible improvements in road safety outcomes. By embedding the monitoring of assumptions within control loops, such as temporary traffic management at work zone, in-vehicle eCall communication, and trauma triage with correct and direct conveyance, the approach addresses the persistent gap between intended safety performance and operational reality.
Looking at the first case study presented in this paper about the eCall delivery\cref{tab:assumptions-ecall}, regular supervision of assumptions related to availability of the the function, the completeness of data, and communication latency ensures that emergency notifications are delivered reliably and and within the acceptable time-frames. This assumption is significant given the time-sensitive nature of post-crash interventions. where a reduction in emergency response time is directly associated with improved survivability. and highlighting technical and procedural assumptions, will supports the early identification of weakness within the emergency communication chain.
the second case study presented, and which is related to temporary traffic management\cref{tab:assumptions- Temporary Traffic Management (Work-Zones)}, monitoring assumptions related to the correct and timely deployment of temporary safety measure, ensures that any non-conformities or deviation from approved plan will be detected and addresses in real time. this enables the prompt mitigation of work zone collisions, thereby reducing both the likelihood and severity of such events. 
Trauma triage\cref{tab:assumptions-Destination compliance}, is an essential factors for increasing survivability rates, the accuracy of triage protocols creates a transparent feedback loop that facilitates continuous improvement.


In this context, of this case study on road safety, the modelled control structure provides complete visibility of responsibilities and the interactions.
This transparency exposes gaps, communication delays and mismatches in the control authority path. By linking assumptions to specific control loops, the framework enables precise traceability\cref{LI_CA}: when assumptions to specific control loops, the framework enables precise traceability: when an assumption fails, it becomes possible to identify which stakeholder is responsible, where accountability lies and what corrective coordination is needed. This supports dynamic system adaptation, ensuring that safety assurance evolves in step with system complexity and operational change. This addressed the \textbf{RQ2} (How can systems engineering incorporating systems thinking be effectively combined to enable an injury reduction and an increase in survival rates?) as combining systems engineering with systems thinking enables an actionable pathway from indicator to intervention.
Overall, this study contributes to achieving its objective, reducing road fatalities and improving survival outcomes by:
\begin{itemize}
    \item Revealing hidden dependencies between stakeholders, presumed assumptions, that influence safety performance.
    \item Identifying vulnerabilities in the control loop where an invalid assumption reduces safety effectiveness.
    \item making implicit system assumption explicit, measurable and traceable.
    
\end{itemize}

The European Commission established eight road safety KPIs \cite{EuropeanCommission2022KPIs}, designed for policy benchmarking during 2021--2023. These KPIs provide indicators on critical risk factors and aim to improve road safety. After analysing the control structure and identifying the control actions (CAs) and feedback (FB), a gap analysis was conducted to determine which elements were already covered by the pre-existing KPIs. The CAs and FB not mapped to any existing KPI were assigned to newly defined categories, such as:

\begin{itemize}
    \item Lifecycle Governance
    \item Organisational Competence \& Role Assurance
    \item Data Integrity
    \item Cyber Safety Assurance
    \item Emergency Response Chain
    \item Infrastructure Operations and Temporary Traffic Management
    \item Fleet Operations \& Compliance
    \item Funding \& Strategic
\end{itemize}
The approach's contribution integrates both KPI and non-KPI assumptions across the entire lifecycle, supporting reductions in fatal and serious injury risk by monitoring assumptions such as the reliability of eCall delivery, the correctness of triage for major trauma patients, and compliance rates. This enables the early identification of invalid assumptions and system degradation in underlying processes, allowing the corresponding stakeholders to intervene promptly. Such monitoring improves system responsiveness and post-crash care, which are recognised as key factors influencing road safety.
Beyond these two factors, the approach enhances the safety of the operational environment by monitoring work zones, traffic management compliance, and temporary safety measures. This contributes to reducing collisions and the exposure of roadside workers, thereby lowering the severity of fatalities.
By transforming the static control structure into a dynamic monitoring network, our approach increases road safety. It enables the timely detection of assumption drift, communication errors, and latent control weaknesses that could lead to severe accidents. Consequently, the approach contributes to improving post-crash survival through faster emergency response and harmonised trauma care, while ensuring compliance with safety standards and recommendations across infrastructure and operations. This supports the strategic objective of \textit{Vision Zero} and aligns with the European Union’s 2030 targets for halving road deaths and serious injuries by ensuring that existing performance information is regularly monitored, acted upon, and that corrective actions are initiated whenever deviations occur.

\subsection*{Implication for safe system practice}
From a safe system perspective, The approach impact several principles:
\begin{itemize}
    \item Proactive risk management: Continuous monitoring of assumptions, identification and mitigation of drift of leading indicators.
    \item Shared responsibility: Explicit assignment of indicators and corrective actions to specific stakeholders supports accountability.
    \item Closed loop feedback: Ensures audit and inspection findings are addressed and closed.
    \item Integration across the safety chain: Expanding from traditional KPI and include telematic data, compliance and conformity ensures end to end control of all actors of road safety.
\end{itemize}

\subsection*{Limitation and Future work}
While the framework strengthens assurance, it depends on accurate, timely data, effective implementation, and robust management of the large number of assumptions and corrective actions. Future work should examine integrating automated data collection, refining context-specific leading indicators, and developing a prioritisation approach to manage the voluminous data.

\section{Conclusion}

In conclusion, this study demonstrates that traditional approaches to road safety largely reliant on retrospective crash data and focused on driver behaviour are insufficient to address the persistent stagnation in safety performance. Road safety must instead be recognised as a complex socio-technical system, shaped by interactions across multiple organisational levels, policies, regulatory.

By integrating systems thinking with systems engineering principles, this paper introduces a proactive systems-based safety approach that extends beyond reductionist analysis. Using a systems-theoretic model, a road safety control
structure for England was developed and applied to identify and monitor critical assumptions embedded within system design and decision-making processes. The results show that continuous monitoring of these assumptions
is essential for maintaining safe operation and preventing system drift into hazardous states.
This approach advances road safety practice by shifting the focus from reactive analysis to continuous safety assurance.

The results shows that road safety is a shared responsibility distributed across all stakeholders. survival outcomes after collision, for example, depend not only on paramedics but also on accurate emergency dispatch, safe road access design, and traffic management support. Likewise, speed compliance is shaped not solely by driver behaviour but by the entire road ecosystem's clarity and reliability of roadside systems. These insights emphasise the necessity of integrated safety governance and collaborative safety management across the entire corresponding stakeholder.
Also presumed assumptions are subject to be breached when a system is in evolving environment, which may lead to more severe fatalities which highlight the need for continuous monitoring of these assumptions.
It enables early detection of emerging risks and supports timely corrective interventions to preserve safe functionality. The findings reinforce the need for a system-wide, lifecycle-oriented strategy for improving road safety and demonstrate the potential of systems engineering as a foundational discipline for future road safety research and policy development.


\section{Acknowledgement}
The work presented in this paper has been carried under the UKRI Future Leaders Fellowship (Grant MR/S035176/1). The authors would like to thank the WMG Centre of HVM Catapult and WMG, University of Warwick, UK, for providing the necessary infrastructure for conducting this study. WMG hosts one of the seven centres that together comprise the High Value Manufacturing Catapult in the UK.

\bibliographystyle{unsrtnat}
\bibliography{library}

\end{document}